\documentclass[12pt]{iopart}
\usepackage{color}
\def\bra#1{\langle #1|}
\def\ket#1{|#1\rangle}

\usepackage{graphicx}

\begin{document}
\title[Entanglement entropy after a partial projective measurement in  CFT: exact results]
{Entanglement entropy after a partial projective measurement in $1+1$ dimensional conformal field theories: exact results }

\author{M.~A.~Rajabpour}
\address{ Instituto de F\'isica, Universidade Federal Fluminense, Av. Gal. Milton Tavares de Souza s/n, Gragoat\'a, 24210-346, Niter\'oi, RJ, Brazil}

\date{\today{}}

\begin{abstract}

We calculate analytically the R\'enyi bipartite entanglement entropy $S_{\alpha}$ of the ground state of $1+1$ dimensional conformal field theories (CFT)
after performing a projective measurement in a part of the system. We show that the entanglement entropy in this setup
is dependent on the central charge and the operator content of the system.
When  the measurement region $A$ separates the two parts $B$ and $\bar{B}$, the entanglement entropy between $B$ and $\bar{B}$
decreases like a power-law  with respect to the characteristic distance between the
two regions with
an exponent which is dependent
on the rank $\alpha$ of the R\'enyi  entanglement entropy and the smallest scaling dimension present in the system. We check our findings by making numerical calculations on the
Klein-Gordon
field theory (coupled harmonic oscillators) after fixing the position (partial measurement) of some of the oscillators. We also
comment on the post-measurement entanglement entropy in the massive quantum field theories.
\end{abstract}
\maketitle
\section{Introduction}

Entanglement entropy has been playing an important role both in the high-energy and the condensed matter  physics for many years. One of 
the interesting aspects of the bipartite entanglement entropy of the critical systems in $1+1$ dimensions is its logarithmic dependence to the 
subsystem size with a coefficient which is dependent on the central charge of the underlying CFT \cite{Holzey1994,Calabrese2009}. 
Accordingly 
bipartite entanglement entropy has been used to get information about the underlying quantum field theory of the system. To get more information about
the quantum field theory one can either  calculate the entanglement entropy of two-disjoint intervals \cite{one dimension MI  numeric, one dimension MI theory}
or the entanglement entropy of the excited states of the system \cite{Chico}. The above quantities are usually dependent on the full operator content of the system
and so can fix the CFT. They are also extremely useful in numerical calculations to find the universality class of
a fixed point in the studies of  quantum phase transitions. Analytical and numerical aspects of bipartite entanglement entropy of
$1+1$ dimensional critical systems is now well understood. However, little is known about
entanglement entropy of tripartite systems. One of the quantities that have been investigated in this regard is the  
quantum entanglement 
negativity studied in the context of CFT in \cite{negativity}, see also references therein. It is a very useful quantity when
one needs to calculate the bipartite entanglement entropy of two domains after tracing out the third party.

One natural question that arises in the context of tripartite systems  regards the behavior of entanglement entropy after partial
projective measurement in the system. 
In other words, consider the ground state of a many body system and make a local projective measurement in a subsystem and then look at the bipartite 
entanglement entropy of
the remaining part. 
In general, an answer to such kind of questions can be dependent on the basis that one chooses to perform the measurement and of course
also to the outcome of the measurement. For example,
by considering a simple interacting spin system one can immediately realize that there are infinite possibilities to perform the measurement and so naturally
the general solution to such kind of problems looks difficult. However, recently it was possible to show \cite{Rajabpour2015b} that
for quantum  chains that can be described by CFT there are some natural  bases \cite{AR2013,AR2015}, see also \cite{Stephan2013,Stephan2014,Najafi} 
that do not destroy the CFT structure of the system and so
one can calculate analytically the bipartite entanglement entropy.  
 When the 
system under study is a critical system  the measurements in the those bases lead to a boundary conformal field theory in the Euclidean space. 
For this reason, we  call them "conformal" bases. However, one should notice that in some cases 
even if one does the measurement in the mentioned natural bases  
the outcome of the measurement can be a configuration which from the renormalization group point of view might not 
flow to a conformal boundary condition, see for example \cite{Najafi}.
In \cite{Rajabpour2015b} it was shown that the ground state
after partial measurement still follows the logarithmic
law but the formulas change slightly. In this work, we want to generalize the same idea and investigate the post-measurement entanglement
entropy in $1+1$ dimensional CFTs in different situations. Using the method first introduced in \cite{Holzey1994} and further elaborated  in
\cite{Cardy2015} we find exact solutions for the post-measurement
entanglement entropy in different geometries. Using this method, we will derive some general formulas that the well-known
results of the entanglement entropy of a sub-region will be just special cases. We will also find some results regarding the entanglement 
entropy of disconnected regions.
Then we check numerically our findings  in
the case of the Klein-Gordon field theory by using the projective measurement of a position of the oscillators. We will also
provide some numerical results regarding post-measurement entanglement entropy in the massive field theories. It is worth mentioning that similar
 questions as we are interested here are already discussed in the context of localizable entanglement entropy \cite{localizable}.
In the next section, we first describe the set up of the problem  and define our system of interest.
Then in section ~3, we find exact solutions for the entanglement entropy of the tripartite system.
In section ~4, we study the finite size effect. In section ~5 by numerical means we  discuss 
the post measurement bipartite entanglement entropy in $1+1$ dimensional massless and massive Klein-Gordon
field theory.  Finally in section ~6 we summarize our results.

\section{Setup and definitions}

Consider  a quantum system  and divide the system
into two subsystems $D$ and $\bar{D}$. The von Neumann entanglement entropy of $D$ with respect to $\bar{D}$
is defined by the following formula:
\begin{eqnarray}\label{von Neumann}
S=-\tr\rho_D\ln\rho_D,
\end{eqnarray}
where $\rho_D$ is the reduced density matrix of the subsystem $D$. A simple generalization of the von Neumann
entropy which is also a measure of entanglement is called R\'enyi entropy and is defined as
\begin{eqnarray}\label{Renyi}
S_{\alpha}=\frac{1}{1-\alpha}\ln\tr\rho_D^{\alpha}.
\end{eqnarray}
The limit $\alpha\to 1$ gives back the von Neumann entropy. What we are interested 
in  is the following: consider a generic quantum system in its ground state and make local projective measurements 
in a sub-region $A$ of this system. After measurement the system  collapses to a wave function in which the subsystem $A$ is
disentangled from its complement $\bar{A}$. However, the subsystem $\bar{A}$ has a wave function which is highly entangled. Now divide the subsystem
$\bar{A}$ to two subsystems $B$ and $\bar{B}$ in a way that $\bar{A}=B\cup\bar{B}$. What we are interested in is the entanglement entropy between $B$ and $\bar{B}$. This setup was first 
studied in \cite{Rajabpour2015b} for the  $1+1$ dimensional CFT's when $B$ and $\bar{B}$ are connected and the measurement
is done in the conformal basis. Here we study a more general case when $A$ is not necessarily a simply connected domain. The most
general case that we would like to study is shown in the Figure ~1. It will be soon clear how this setup contains 
also the cases that have already been  discussed in the literature before.

{\it{Notation:}} In this work $s_1(s_2)$  and $l$ represent the characteristic  sizes of $A$ and $B$.
\begin{figure} [htb] \label{fig1}
\center
\includegraphics[width=0.6\textwidth]{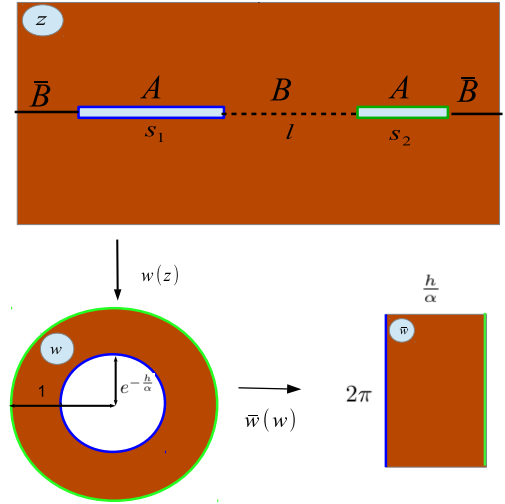}
\caption{(Color online) Mapping between different regions. The whole plane with two slits $A$ and branch cut (dashed line) on $B$ can be mapped to
annulus  by the conformal map $w_{\alpha}(z)$. Then one can go to a cylinder with the conformal map $\bar{w}(w)=\ln w$. } 
\end{figure}

\section{Analytical calculation of the post-measurement bipartite entanglement entropy}

To calculate the entanglement entropy we generalize the method of \cite{Holzey1994}, see also \cite{Cardy2015}.  It is quite well-known that
the R\'enyi 
entanglement entropy of a generic system in $d+1$ dimensions can be written as the ratio of the partition functions in $d+1$ dimensions
as \cite{Holzey1994,Calabrese2009}:
\begin{eqnarray}\label{Reimann surfaces}
S_{\alpha}=\frac{1}{1-\alpha}\ln\frac{Z_{\alpha}}{Z_1^{\alpha}},
\end{eqnarray}
where $Z_{\alpha}$ is the partition function of the system on $\alpha$-sheeted Riemann surfaces. For example, in two dimensions one just needs to consider the
partition function of the system in the presence of a branch cut along the region $B$ that we would like to calculate
its entanglement entropy. The above results are valid independent
of the boundary conditions and the dimensionality of the system. Consider now the setup that we provided in the previous section. After
performing projective measurement in part $A$ of the system that part will be decoupled from the rest, so to calculate
the entanglement entropy of $B$ with respect to $\bar{B}$ in the Euclidean language we just need to calculate the 
partition function of $\alpha$-sheeted Riemann 
surfaces with slits on the region $A$. In other words, the system of interest is as Figure ~1 with branch cut on $B$ and slits on $A$. Different
Riemann surfaces are connected through the branch cut. In general calculating the above quantity is a very difficult task, however, when
the system is at the critical point and the boundaries are conformally invariant as we will show one can evaluate $S_{\alpha}$ analytically. 
As we stated in the introduction the boundary at $A$ is a conformal one if one makes the projective measurement in the conformal" 
bases \cite{AR2013,Rajabpour2015b}.
For example, for the free bosonic system (coupled harmonic oscillators) the field $\phi$ 
measurement corresponds to Dirichlet boundary condition which is a conformal boundary. It is worth mentioning that since the region $A$
is a bipartite domain in principle one can have different conformal boundary conditions on the two parts of the domain $A$
and this can affect the result of the entanglement entropy.  Although this effect does not change the leading term, we will briefly
comment about
this phenomena later, for a recent discussion see \cite{Cardy2015}.

Instead of directly working with the partition function it is much better to work with the free energy defined as $\mathcal{F}_{\alpha}=-\ln Z_{\alpha}$.
To calculate the free energy on the Riemann surface with slits we first map the surface to the annulus using the conformal map
$w(z)$. As we will show the conformal map that we use to map 
the Riemann surface
to the annulus contributes to the free energy of the Riemann surface. We call this contribution the geometric part of the free energy 
$\mathcal{F}^{geom}_{\alpha}=-\ln Z^{geom}_{\alpha}$, where $Z^{geom}_{\alpha}$ is the corresponding partition function. After the conformal 
map one is left with the free energy on the annulus
which we call $\mathcal{F}^{annn}_{\alpha}=-\ln Z^{annu}_{\alpha}$, where $Z^{annu}_{\alpha}$ is the corresponding partition function
on the annulus.  The  free energy on the Riemann surface can be calculated explicitly as follows: 
 the change in the free energy of the system after a small horizontal displacement 
 $\delta l$ of one of  the slits is given by the standard result 
of CFT:
\begin{eqnarray}\label{deformation}
\delta \mathcal{F}_{\alpha}=-\frac{\delta l}{2\pi i}\oint_{\partial S_2} \langle T(z)\rangle dz+ c.c.,
\end{eqnarray}
where the integral is along a contour  $\partial S_2$ 
surrounding one of the slits (here the right one).  Under conformal map $f$ the energy momentum tensor transforms according
to $T(z)=(\partial_zf)^2T(f)+(c/12)\{f,z\}$, where $\{f,z\}=\frac{ f'''}{f'}-\frac{3}{2}(\frac{f''}{f'})^2$ is the Schwarzian derivative. Consequently
to calculate the $\delta \mathcal{F}_{\alpha}$ one needs to map the system
to a geometry which the expectation value of the energy-momentum tensor is known. On the cylinder we have
$2\langle T(\bar{w})\rangle =\frac{\delta \mathcal{F}_{\alpha}^{cyl}}{\delta h_{\alpha}}$, where $h_{\alpha}$ is the length of the cylinder. In addition the free energy on the cylinder is
related to the free energy on the annulus by the standard formula of CFT, i.e. $\mathcal{F}^{cyl}_{\alpha}=\mathcal{F}^{ann}_{\alpha}-\frac{c}{12}h_{\alpha}$. To calculate
the expectation value of the energy-momentum tensor on the Riemann surface we first map the system to the annulus 
by the conformal map $w_{\alpha}(z)$ and then to the cylinder by the conformal map $\bar{w}(w)=\ln w_{\alpha}$.
Finally using the above formulas we arrive to the result
\begin{eqnarray}\label{energy-momentum transformation}
T(z)=\frac{1}{2}(\partial_z\bar w(z))^2\frac{\delta\mathcal{F}^{ann}_{\alpha}}{\delta h_{\alpha}}+\frac{c}{12}\{w_{\alpha},z\}.
\end{eqnarray}
 Putting the above formula in (\ref{deformation}) and using the fact that 
$\frac{\delta h_{\alpha}}{\delta l}=\frac{i}{2\pi}\oint_{\partial S_2}(\partial_z\bar{w})^2$, see for example \cite{Kardar}, one  arrives
to the following result
\begin{eqnarray}\label{main result1}
\mathcal{F}_{\alpha}=\mathcal{F}^{geom}_{\alpha}+\mathcal{F}^{annn}_{\alpha}
\end{eqnarray}
with
\begin{eqnarray}\label{main result2}
\frac{\delta\mathcal{F}^{geom}_{\alpha}}{\delta l}= \frac{ic}{12\pi}\oint_{\partial S_2}\{w_{\alpha},z\}dz,
\end{eqnarray}

The geometric part of the 
free energy is  dependent on only the central charge of the system. However, annulus part of the free energy 
 is dependent on the full operator content of the CFT \cite{Cardy2004}.
Similar calculations as above are already 
done in the context of Casimir force  \cite{Machta,Kardar} and also in the studies of formation probabilities in \cite{RajabCasimir}, for other related works
see \cite{Maghrebi}. Note that the free energy of the Riemann surface with two slits is just like the Casimir energy of two slits on
the Riemann surface. From this perspective the above calculation provides a connection between post-measurement entanglement entropy
and the Casimir energy of two slits.

The conformal map from the plane with two slits on a line with lengths $s_1$ and $s_2$ and a branch cut with the length
$l$ to annulus with the inner 
and outer radiuses $r=e^{-h_{\alpha}}$ and $r=1$ with $h_{\alpha}=\frac{h}{\alpha}$, see Figure ~1, has been derived in the appendix. It has the following form:

\begin{eqnarray}\label{conformal map1}
w_{\alpha}(z)&=&\Big{(}e^{-\frac{h}{2}}e^{h \frac{\mbox{sn}^{-1}(\tilde{z},k^2)}{2\mathcal{K}(k^2)}}\Big{)}^{\frac{1}{\alpha}},\\
h&=&2\pi\frac{\mathcal{K}(k^2)}{\mathcal{K}(1-k^2)},
\end{eqnarray}
where $\mathcal{K}$ and $\mbox{sn}^{-1}$ are
the elliptic  and inverse Jacobi functions \footnote{Note that in all of the formulas we adopt
the Mathematica convention for all the elliptic functions.} respectively and
\begin{eqnarray}\label{conformal map1 details}
\tilde{z}&=&\frac{2a}{k}\frac{z}{bz+1}-\frac{1}{k},\nonumber\\
a&=&\sqrt{\frac{s_2(s_2+l)}{s_1(s_1+l)}}\frac{1}{l+s_1+s_2},\nonumber\\
b&=&\frac{\sqrt{s_1s_2(l+s_1)(l+s_2)}-s_2(l+s_1)}{(l+s_1)(s_1s_2-\sqrt{s_1s_2(l+s_1)(l+s_2)})},\nonumber\\
\end{eqnarray}
with the parameter $k$ given by
\begin{eqnarray}\label{conformal map1 details}
k=1+2\frac{s_1s_2-\sqrt{s_1s_2(l+s_1)(l+s_2)}}{l(l+s_1+s_2)}.
\end{eqnarray}

 The Schwarzian derivative has poles at $z=0,s_1,l+s_1$
and $z=s_1+l+s_2$. Since the contour integral is around the last two poles one can sum over the residue of them and find

\begin{eqnarray}\label{geometric general}
\frac{\delta \ln Z^{geom}_{\alpha}}{\delta l}=
\frac{c\alpha}{6}\frac{\Big{(}(-2a+b)^2-b^2k\Big{)}\Big{(}2\pi^2-(1+k(6+k))\alpha^2\mathcal{K}^2(1-k^2)\Big{)}}{16ak(1+k)\alpha^2\mathcal{K}^2(1-k^2)}.\hspace{0.5cm}
\end{eqnarray}
Note that an extra $\alpha$ appears because we integrate over $\alpha$-sheeted surface. Unfortunately, the $\ln Z^{geom}_{\alpha}$ can not be calculated analytically in the most general cases and so 
one needs to rely either on numerical calculations or study it in special cases. 
We will provide later some analytical results in special limits.

The annulus part of the partition function can be written in two equivalent forms as follows\cite{Cardy2004}:
\begin{eqnarray}\label{annulus part1}
\ln Z^{annu}_{\alpha}&=&\ln [q_{\alpha}^{-c/24}(1+\sum_jn_jq_{\alpha}^{\Delta_j})]-c\frac{h}{12\alpha},\\
\label{annulus part2}
\ln Z^{annu}_{\alpha}&=&\ln [\tilde{q}_{\alpha}^{-c/24}(b_0^2+\sum_jb^2_j\tilde{q}_{\alpha}^{\Delta_j})]-c\frac{h}{12\alpha},
\end{eqnarray}
where $n_j$ and $b_j$ are numbers and $\Delta_j$ in the first formula is the boundary scaling dimension and in the second formula is the
bulk scaling dimension. Finally $q_{\alpha}$ and $\tilde{q}_{\alpha}$ are defined as
\begin{eqnarray}\label{q tilde q}
q_{\alpha}=e^{-\pi\frac{2\pi\alpha}{h}},\hspace{1cm}\tilde{q}_{\alpha}=e^{-\frac{2h}{\alpha}}.
\end{eqnarray}
The first parts of the equations (\ref{annulus part1}) and (\ref{annulus part2}) are equal to $\ln Z^{cyl}_{\alpha}$ which
is equal to the logarithm of the partition function on the finite cylinder and the term $c\frac{h}{12\alpha}$ takes us back to
the annulus. Having the above formulas one can now calculate the entanglement entropy, however, before going to the details of the calculations
it is worth mentioning that one can write the equation (\ref{Reimann surfaces}) with respect to the free energy as
\begin{eqnarray}\label{free energy}
S_{\alpha}=\frac{1}{\alpha-1}\Big{(}\mathcal{F}^{ann}_{\alpha}+\mathcal{F}^{geom}_{\alpha}-\alpha(\mathcal{F}^{ann}_{1}+\mathcal{F}^{geom}_1)\Big{)}.
\end{eqnarray}
Having equipped with all the necessary formulas  we now apply them to the most interesting cases:

\subsection{ $s_1=s_2=s\ll l$:}  

This is the familiar case of the entanglement entropy of the subsystem with the length $l$ with respect to the rest
and can be studied independent of the projective measurement setup \cite{Cardy2015}. In this case $k=  \frac{l}{2s+l}\to 1$
and then $h\to -2\ln\frac{2s}{l}\to \infty$. In this limit it is much better to work with $\tilde{q}=(\frac{2s}{l})^{\frac{4}{\alpha}}+...$ which is  small
as far as $\alpha$ is not too big. When $s_1=s_2=s$
the geometric contribution can be written as
\begin{eqnarray}\label{geometric s1 s2 equal}
\frac{\delta \ln Z^{geom}_{\alpha}}{\delta l}=\frac{c\alpha}{24}\frac{\pi^2(l+2s)^2-2(2l^2+4ls+s^2)\alpha^2\mathcal{K}^2(\frac{4s(l+s)}{(l+2s)^2})}
{l(l+s)(l+2s)\alpha^2\mathcal{K}^2(\frac{4s(l+s)}{(l+2s)^2})}.
\end{eqnarray}
In the limit  $s\ll l$ we can expand the above formula and get

\begin{eqnarray}\label{expansion of geometric part}
\frac{\delta \ln Z^{geom}_{\alpha}}{\delta l}=\frac{c\alpha}{6}(\frac{1}{\alpha^2}-1)(\frac{1}{l}-\frac{s}{l^2}+\frac{3s^2}{2l^3}-\frac{5s^3}{2l^4})
-\frac{c}{64}\frac{s^4(47-48\alpha^2)}{\alpha l^5}+...,
\end{eqnarray}
where the dots are subleading terms. The first non-zero term for $\alpha=1$ starts from the fifth term $\frac{s^4}{l^5}$. Note that  the $\alpha=1$
is  related to the partition function in the Casimir effect studies \cite{Kardar,RajabCasimir}. Using (\ref{annulus part2}) the annulus part can be written as
\begin{eqnarray}\label{expansion of annulus part}
 \ln Z^{annu}_{\alpha}=2\ln b_0+\frac{b_1^2}{b_0^2}(\frac{2s}{l})^{\frac{4\Delta_1}{\alpha}}+...,
\end{eqnarray}
where $\Delta_1$ is the smallest dimension in the conformal tower.
Putting all the terms together we have
\begin{eqnarray}\label{EE subsreion}
S_{\alpha}=\frac{c}{6}\frac{1+\alpha}{\alpha}\ln \frac{l}{s}+2\ln b_0+\frac{b_1^2}{b_0^2}(\frac{2s}{l})^{\frac{4\Delta_1}{\alpha}}+....
\end{eqnarray}
The first term is the well-known formula of the entanglement entropy of the sub-region \cite{Holzey1994,Calabrese2009}.
The second term is the famous Affleck-Ludwig boundary term \cite{Affleck-Ludwig} studied already in the context of entanglement entropy in \cite{Calabrese2009}.
The third term and the dots are the subleading terms and are dependent 
on the scaling dimensions $\Delta_j$ and are already studied in \cite{CardyCalabreseCorrections}. Note that
the leading term of the entanglement entropy  comes from the geometric part of the partition function and the most important subleading terms come from
the annulus part of the partition function.

\subsection{ $s_2 \ll s_1, l$ :}   

This is the limit which has been discussed already in \cite{Rajabpour2015b}. In this limit we have
\begin{eqnarray}\label{tilde q}
h&=&-\ln\frac{s_2s_1}{2l(l+s_1)}+...,\\
\tilde{q}&=&(\frac{s_2s_1}{2l(l+s_1)})^{2/\alpha}+....
\end{eqnarray}
The equation (\ref{geometric general})  after expansion is
\begin{eqnarray}\label{expansion of geometric part}
\frac{\delta \ln Z^{geom}_{\alpha}}{\delta l}=\frac{c(1-\alpha^2)}{12\alpha}\frac{2l+s_1}{l(l+s_1)}+....
\end{eqnarray}
Putting all together we have
\begin{eqnarray}\label{rajab2015}
S_{\alpha}=\frac{c}{12}(\frac{1+\alpha}{\alpha})\ln\frac{l(l+s_1)}{s_2s_1}+2\ln b_0+\frac{b_1^2}{b_0^2}(\frac{s_2s_1}{2l(l+s_1)})^{2\Delta_1/\alpha}+ ...,
\end{eqnarray}
where the dots are the subleading terms that some of them are $\Delta_j$ dependent. Note that again the leading term
comes from the geometric part of the free energy and the most important subleading terms are the result of the annulus part of the free energy. 
The above equation is the same as the result derived in \cite{Rajabpour2015b}
using twist operator technique.

\subsection{ $l\ll s_1=s_2=s$:}   

In this case we are interested to the entanglement entropy of two regions that are because of the 
projective measurement region $A$ completely disconnected 
and far from each other. In this limit it is much better to work with $q$ which is going to be our small parameter
as far as $\alpha$ is not too small. We have
\begin{eqnarray}\label{ q}
h=\frac{\pi^2}{\ln\frac{8s}{l}}+...,\hspace{0.6cm}q=(\frac{l}{8s})^{2\alpha}+....
\end{eqnarray}
Using the equations (\ref{geometric general}) and (\ref{annulus part1}) one can get 
\begin{eqnarray}\label{expansion of geometric part 2 geom}
\ln Z^{geom}_{\alpha}=\frac{c\alpha}{12}\big{(}\frac{\ln\frac{l}{a}}{2}+\frac{\pi^2}{\alpha^2\ln\frac{8s}{l}}\Big{)}+...,\\
\label{expansion of geometric part 2 annu}
\ln Z^{annu}_{\alpha}=\frac{c\alpha}{12}\big{(}\ln\frac{8s}{l}-\frac{\pi^2}{\alpha^2\ln\frac{8s}{l}}\Big{)}+n_1(\frac{l}{8s})^{2\alpha\Delta_1}+...,
\end{eqnarray}
where we take $a$ as the smallest distance between the two regions which its presence is necessary here to keep the integral
convergent. After summing over all the terms  we  have
\begin{equation}\label{power-law decay}
{S_{\alpha}}\asymp \left\{
\begin{array}{c l}      
    \frac{1}{\alpha-1} (\frac{l}{8s})^{2\alpha\Delta_1}, & \alpha<1\\
        (\frac{l}{8s})^{2\Delta_1}\ln\frac{l}{8s}, &\alpha=1\\
        \frac{\alpha}{\alpha-1} (\frac{l}{8s})^{2\Delta_1}, & \alpha>1,
\end{array}\right.
\end{equation}
where $\Delta_1$ is the smallest boundary scaling dimension in the spectrum of the system. Note that the first two terms of the equation
(\ref{expansion of geometric part 2 annu}) are canceled out by the first two terms of the equation (\ref{expansion of geometric part 2 geom})
in a way that the leading term of $S_{\alpha}$ comes from the annulus part of the free energy.
The first interesting aspect of the above result is that the two regions although not connected to each other 
are highly entangled. We will show in the next section that this is mainly because one can consider the post-measurement wave 
function as the wave function of a non-local Hamiltonian (we call it from now on corresponding post-measurement Hamiltonian) 
with power-law decaying couplings. This property makes the setup useful to prepare power-law entangled disconnected many body systems
in experimental investigations. The second interesting feature is that for large distances if one increases the size of the 
subsystem $l$ the entanglement entropy will increase with a power-law like function with respect 
to the $l$ which again one should trace its reason back to the non-local feature of the corresponding post-measurement Hamiltonian. Finally
note that the decaying exponent is $\alpha$-dependent
which makes the above result distinct from  the entanglement entropy of two disjoint intervals studies in 
\cite{one dimension MI  numeric, one dimension MI theory,higher dimension MI  theory,higher dimension MI numeric}. 
Since for $\alpha_1\leq\alpha_2$ we still have $ S_{\alpha_2}\leq S_{\alpha_1}$ the above result is consistent with the general monotonicity properties
of R\'enyi entanglement entropy. However, in most of the previous studies for large distances or large sizes one get $\frac{S_{\alpha_2}}{S_{\alpha_1}}\to r$ 
where $r>0$  is a constant.
In our case for $\alpha_1<1$ the ratio $r$ is zero for large distances and grows with keeping $s$ fixed and increasing $l$ which means 
that the behavior of the smallest eigenvalues of  the system after projective measurement is very different from the cases without 
projective measurement. Studying the spectrum of the entanglement 
Hamiltonian after projective measurement might shed more light on the understanding of distinct behavior in the $\alpha<1$ regime. 
It is worth mentioning that if $\Delta_1$ is too big then the leading power-law term can come from the
 the geometric  part of the  free energy with an exponent which
is independent of $\alpha$.

\subsection{ Different boundary conditions:} 

In this subsection we comment on the possible effect of
the different boundary conditions on the two slits. When there are two different conditions on the boundaries of the annulus, i.e. $A$ and $B$, the equations
(\ref{annulus part1}) and (\ref{annulus part2}) have the following more general forms \cite{Cardy2004}:
\begin{eqnarray}\label{annulus part3}
\ln Z^{annu}_{\alpha}&=&\ln [q_{\alpha}^{-c/24}(1+\sum_jn^{AB}_jq_{\alpha}^{\Delta_j})]-c\frac{h}{12\alpha},\\
\label{annulus part4}
\ln Z^{annu}_{\alpha}&=&\ln [\tilde{q}_{\alpha}^{-c/24}(b^{A}_{0}b^B_{0}+\sum_jb^{A}_{j}b^B_{j}\tilde{q}_{\alpha}^{\Delta_j})]-c\frac{h}{12\alpha},
\end{eqnarray}
where $n^{AB}_j$ are the non-negative integers and $b^A_{j}=\langle A|j\rangle\rangle$ and $b^B_{j}=\langle\langle j|B\rangle$ with 
$|A(B)\rangle $ and $|j\rangle\rangle$ being Cardy
and Ishibashi states respectively. Different coefficients are related to each other with the formula $n_j^{AB}=\sum_{j'}S_j^{j'}b_j^{A}b_{j'}^B$,
where $S_j^{j'}$ is the element of the modular matrix $S$,
for more details see \cite{Cardy2004}.  As it should be clear now 
the effect of the different boundary conditions on the entanglement entropy is subleading. The first non-zero contribution, for example, in the case of 
$(s_2 \ll s_1,l)$ can be easily calculated by expanding $\tilde{q}$ as before and then putting the result in (\ref{annulus part4})
and finally in (\ref{free energy}). The extra contribution coming from the boundary conditions is 
\begin{eqnarray}\label{ludwig-affleck}
S^{LA}=\ln b_0^A+\ln b_0^B
\end{eqnarray}
This is again the Affleck-Ludwig boundary term \cite{Affleck-Ludwig} studied already in the context of entanglement entropy in \cite{Cardy2004}.

\section{Finite size effect}

In this subsection we extend the results to a periodic system with finite size. The desired conformal map from
an infinite cylinder with two aligned slits to annulus is as follows: First  we introduce the conformal 
map $\tilde{z}(z)$, which takes the system from infinite cylinder with two slits
to the whole plane with two symmetric aligned slits on the real line as we discussed in the appendix. The map is as follows:
\begin{eqnarray}\label{conformal map2 details}
\tilde{z}&=&\frac{e^{2i\pi\frac{z}{L}}+a_0}{b_1e^{2i\pi\frac{z}{L}}+b_0},\\
a_0&=&\frac{e^{2i\pi\frac{s_1}{L}}}{N}\Big{(}1-k-2e^{2i\pi\frac{l+s_1}{L}}+(1+k)e^{2i\pi\frac{l}{L}}\Big{)},\nonumber\\
b_1&=&\frac{-1}{N}\Big{(} (1-k)e^{2i\pi\frac{l+s_1}{L}}+2k-(1+k)e^{2i\pi\frac{s_1}{L}}\Big{)},\nonumber\\
b_0&=&\frac{e^{2\pi\frac{s_1}{L}}}{N}  \Big{(}1-k+2ke^{2i\pi\frac{l+s_1}{L}}-(1+k) e^{2i\pi\frac{l}{L}}\Big{)},\nonumber\\
N&=&-2-e^{2i\pi\frac{l+s_1}{L}}(-1+k)+e^{2i\pi\frac{s_1}{L}}(1+k),\nonumber
\end{eqnarray}
with the $k$ given by
\begin{eqnarray}
 k&=&1+2\frac{\sin[\frac{\pi s_1}{L}]\sin[\frac{\pi s_2}{L}]-\sqrt{\sin[\frac{\pi s_1}{L}]\sin[\frac{\pi s_2}{L}]\sin[\frac{\pi (s_1+l)}{L}]
\sin[\frac{\pi (s_2+l)}{L}]}}{\sin[\frac{\pi l}{L}]\sin[\frac{\pi (l+s_1+s_2)}{L}]}.
\end{eqnarray}
Now again we can use the same conformal map as (\ref{conformal map1}) to map the whole plane with two aligned symmetric slits and a branch cut
to the annulus.
 The Schwarzian derivative and  the integral (\ref{main result2}) over the poles at $z=s_1+l$ and $z=s_1+l+s_2$ can be calculated using Mathematica.
 The final result has  the following form:

\begin{eqnarray}\label{finite size geometric}
\frac{\delta \ln Z^{geom}_{\alpha}}{\delta l}=-i\pi c \frac{P-\alpha^2Q\mathcal{K}^2(1-k^2)}{\alpha R \mathcal{K}^2(1-k^2)}
\end{eqnarray}
with 

\begin{eqnarray}
P=2\pi^2\Big{(}-4k(e^{2\pi i\frac{l+s_1}{L}}-1)+(1+k)^2e^{2\pi i\frac{s_1}{L}}(e^{2\pi i\frac{l}{L}}-1)^2\Big{)},\nonumber\\
Q=(1+6k+k^2)\times\nonumber\\
\Big{(}-2(k-1)^2e^{2\pi i\frac{l+s_1}{L}}-4k-4ke^{4\pi i\frac{l+s_1}{L}}+(1+k)^2e^{2\pi i\frac{s_1}{L}}+(1+k)^2e^{2\pi i\frac{2l+s_1}{L}}\Big{)},\nonumber\\
R=48Lk(1+k)^2(-1+e^{\frac{2i\pi l}{L}})(-1+e^{\frac{2i\pi s_1}{L}})(-1+e^{\frac{2i\pi (s_1+l)}{L}}).\nonumber
\end{eqnarray}
In the limit $L\to \infty$ we are back again to the result (\ref{geometric general}). In the following subsections we will study some special cases of the above formula.

\subsection{ $s_1=s_2=s\ll l$:}

This is the familiar case of the entanglement entropy of a subsystem without projective measurement. As before
the annulus part does not contribute to the leading term. The derivative of the geometric part of the partition function
has the following form
\begin{eqnarray}\label{finite size1}
\frac{\delta \ln Z^{geom}_{\alpha}}{\delta l}=
\frac{\pi(1-\alpha^2)}{6\alpha }\frac{\cot[\frac{\pi l}{L}]}{L}-\frac{\pi^2(1-\alpha^2)}{6\alpha }\frac{s}{L^2\sin^2[\frac{\pi l}{L}]}+...,
\end{eqnarray}
Where the dots are subleading terms. In this case the annulus part of the free energy does not have any contribution
in the leading order. After integration of the equation (\ref{finite size1}) one can easily find
\begin{eqnarray}\label{EE subsreion finite size}
S_{\alpha}=\frac{1}{6}\frac{1+\alpha}{\alpha}\ln \Big{(}\frac{L}{\pi s}\sin[\frac{\pi l}{L}]\Big{)}+...,
\end{eqnarray}
which is the well-known formula of the entanglement entropy of the subregion \cite{Calabrese2009}.

\subsection{ $s_2\ll s_1, l$:}

In this case as far as $\alpha$ is not too big the small parameter is again $\tilde{q}$  and the annulus
part does not play any role in the leading level. After expansion of the formula
(\ref{finite size geometric}) with respect to $s_2$ and  we have

\begin{eqnarray}\label{finite size2}
\frac{\delta \ln Z^{geom}_{\alpha}}{\delta l}=
\frac{c(1-\alpha^2)}{12\alpha}\frac{\pi\sin[\frac{\pi(2l+s_1)}{L}]}{L\sin[\frac{\pi l}{L}]\sin[\frac{\pi(l+s_1)}{L}]}+...
\end{eqnarray}
The entanglement entropy now can be calculated easily after integration of the above formula. the leading term has the following form:

\begin{eqnarray}\label{SB for PBC}
S_{\alpha}=\frac{c}{12}(1+\frac{1}{\alpha})\ln \Big{(}\frac{L}{\pi}\frac{\sin\frac{\pi}{L}(l+s_1)\sin\frac{\pi}{L}l}{s_2\sin\frac{\pi}{L}s_1}\Big{)}+...,
\end{eqnarray}
which is the result of \cite{Rajabpour2015b}. Note that again the annulus part does not have any contribution in the leading term.

\subsection{ $l \ll s_2=s_1=s$:}

In this limit as far as $\alpha$ is not too small  we have
\begin{eqnarray}\label{h and q PBC}
h=\frac{-\pi^2}{2\ln\frac{\pi l}{4L}}+...,\hspace{0.5cm}q=(\frac{\pi l}{4L})^{4\alpha}+....
\end{eqnarray}
Then after using the equations (\ref{annulus part1}) and (\ref{finite size geometric}) we have
\begin{equation}\label{power-law decay 2}
{S_{\alpha}}=\left\{
\begin{array}{c l}      
    \frac{1}{\alpha-1} (\frac{\pi l}{4L})^{4\alpha\Delta_1}, & \alpha<1\\
        (\frac{\pi l}{4L})^{4\Delta_1}\ln\frac{\pi l}{4L}, &\alpha=1\\
        \frac{\alpha}{\alpha-1} (\frac{\pi l}{4L})^{4\Delta_1}, & \alpha>1,
\end{array}\right.
\end{equation}
Two comments are in order here.  First similar to the derivation of  the equation (\ref{power-law decay})
the leading term comes from the annulus part of the free energy. Second 
the exponents here are two times bigger than the ones we discussed before in the equation (\ref{power-law decay}).

\section{Numerical results: Klein-Gordon free field theory}

In this section, we verify the exact formulas derived in the last section by performing numerical calculations on the free bosonic system.
The Klein-Gordon field theory is
defined by the following action: 
\begin{eqnarray} \label{Klein-Gordon free field theory}
\frac{1}{2}\int \lbrace(\partial\phi(x))^2+{m}^{2}\phi^2(x)\rbrace dx.
\end{eqnarray}
The procedure that we follow is based on a well-known technique. We first discretize the field theory and get some coupled harmonic oscillators 
and then use the correlation matrices as first discovered in \cite{Bombelli1986}, see also \cite{Srednicki} and \cite{Casini2009a}, to calculate
the entanglement entropy after projective measurement. The measurement that we are interested in is the field $\phi(x)$
itself which is the most natural basis one can start with. Since fixing $\phi(x)$ is just a Dirichlet boundary condition if one works with a 
massless system
the conformal structure will be still preserved. The upcoming results are also valid  if one
does the measurements in the momentum basis (related to Neumann boundary condition) but not a generic basis that diagonalizes an 
arbitrary field.

Consider the Hamiltonian of $L$-coupled harmonic 
oscillators, with coordinates $\phi_1,\ldots,\phi_L$ and conjugated momenta 
$\pi_1,\ldots,\pi_L$:
\begin{figure} [htb] \label{fig02}
\centering
\includegraphics[width=0.4\textwidth,angle =-90]{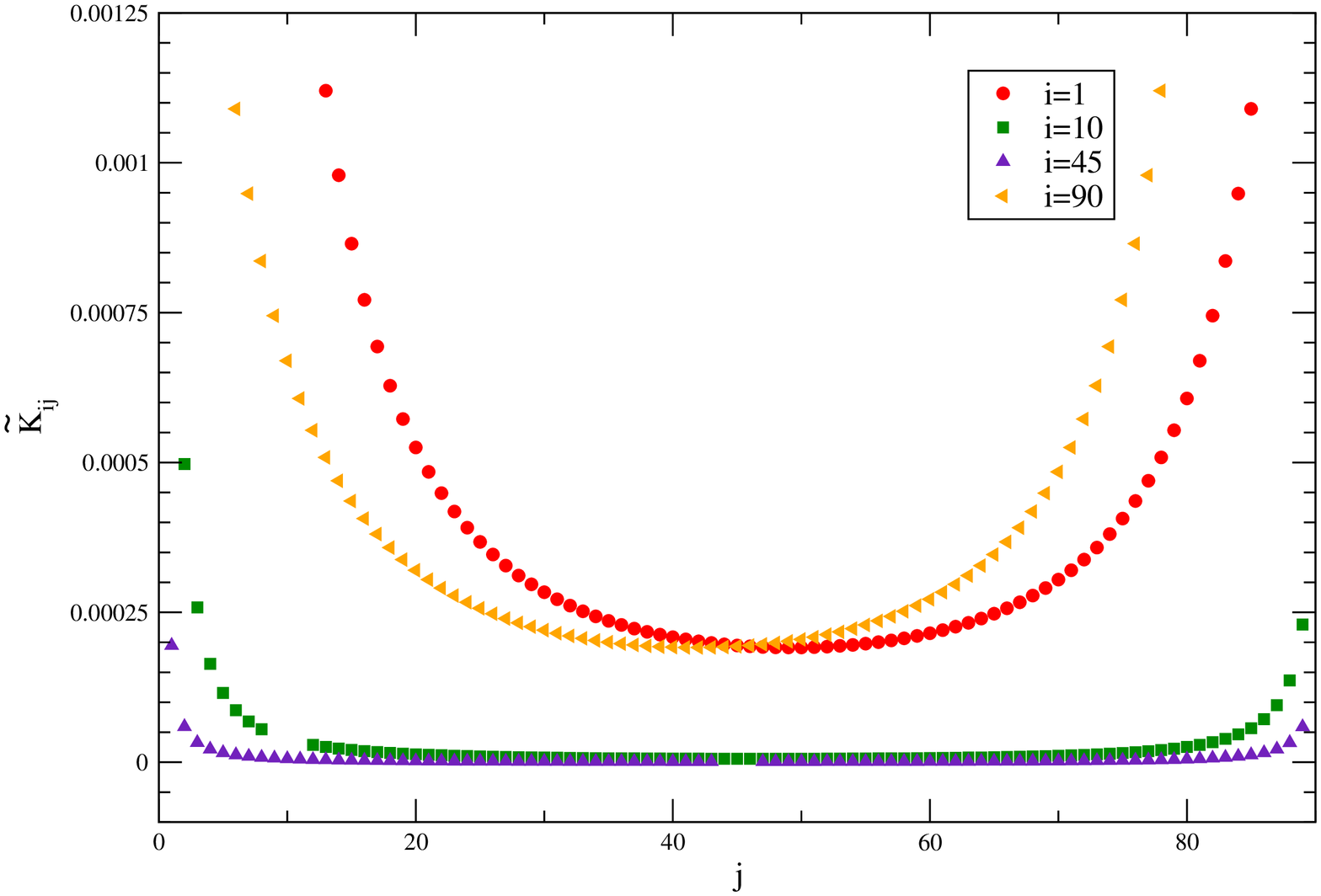}

\includegraphics[width=0.4\textwidth,angle =-90]{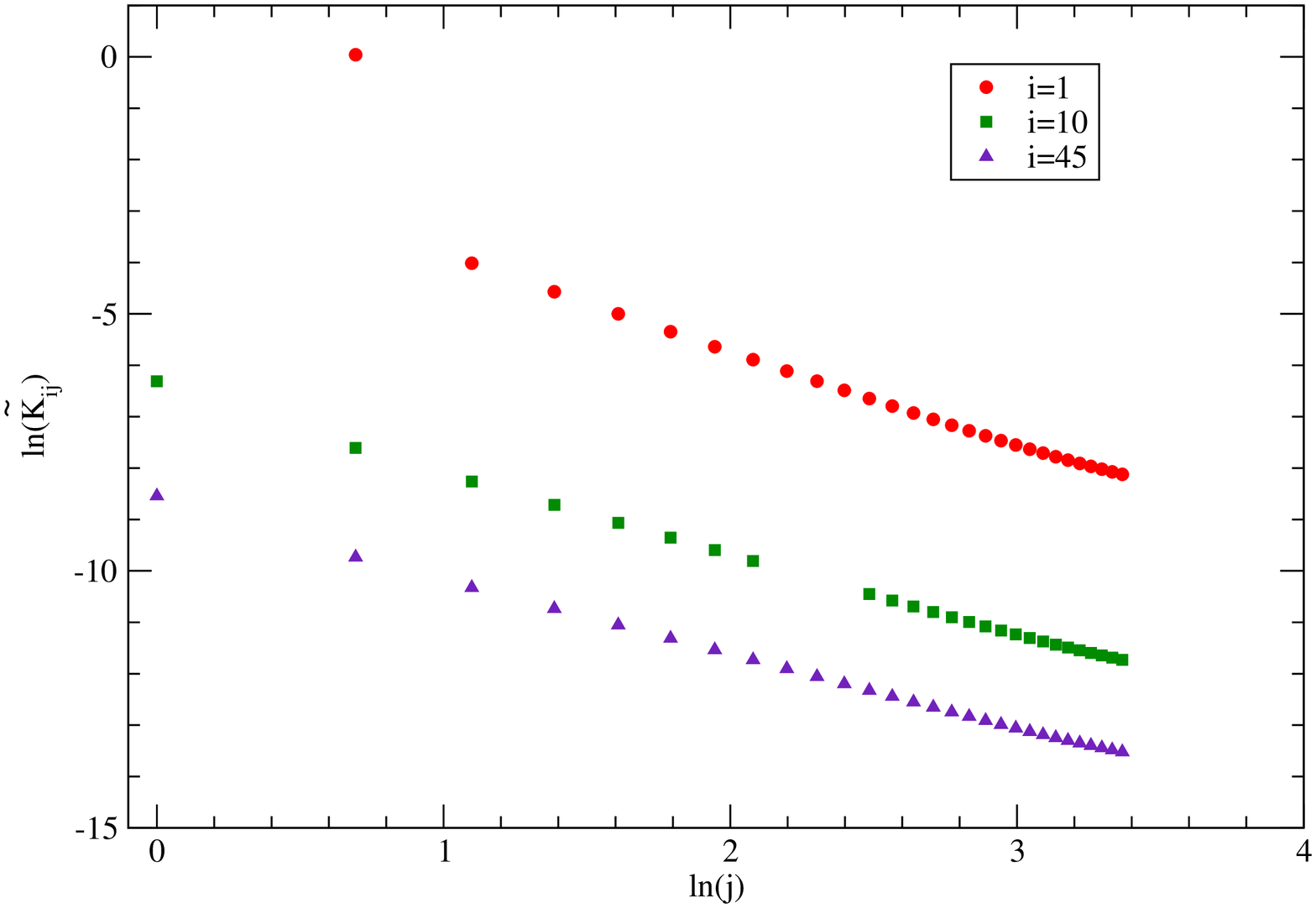}

\includegraphics[width=0.4\textwidth,angle =-90]{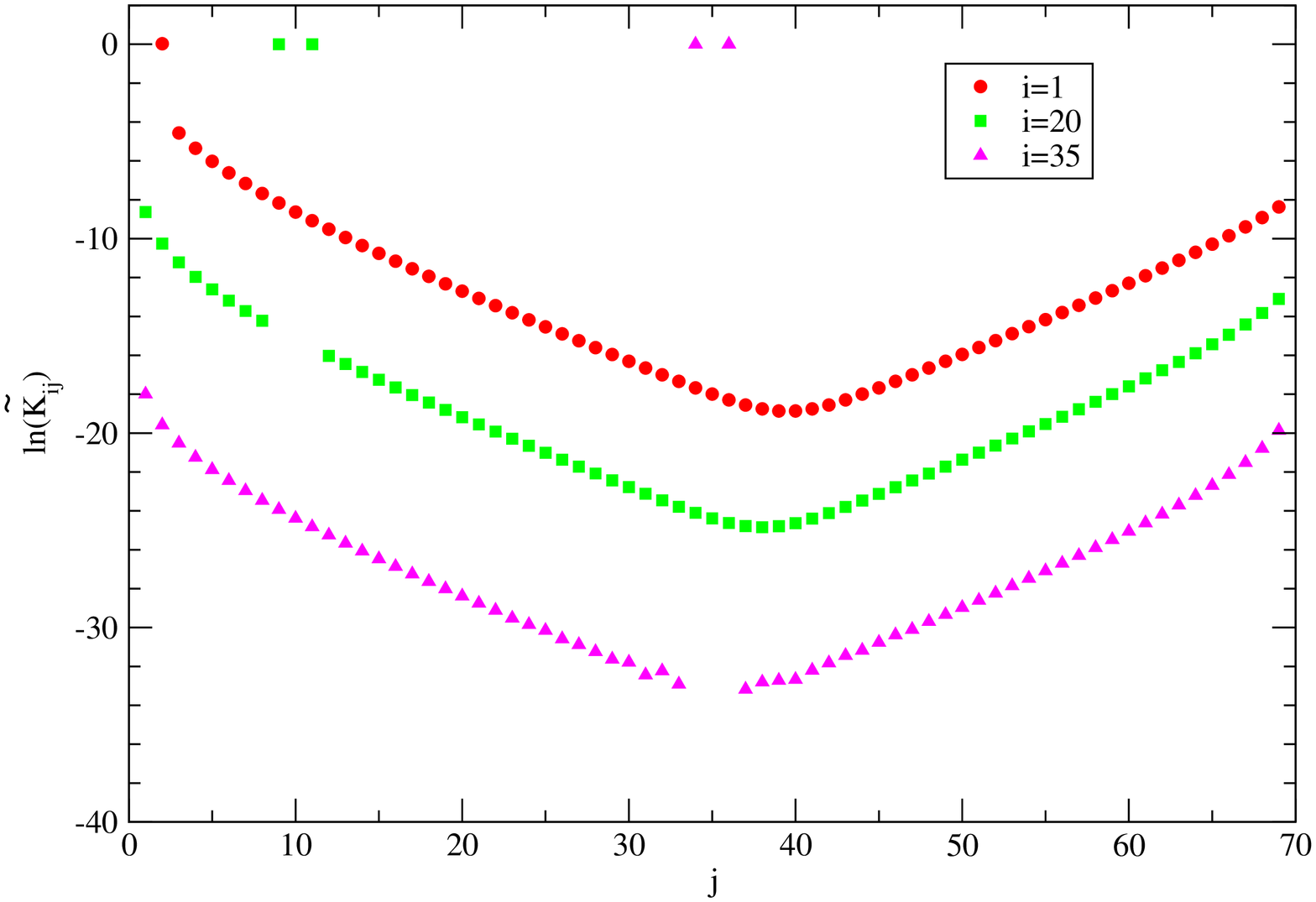}
\caption{(Color online) Top: $\tilde{K}_{ij}$ with respect to $j$ after performing measurements on $10$ contiguous sites for discrete Laplacian with $L=100$.
Middle: The same quantity in the log-log plot up to $j=30$.
Bottom: $\ln\tilde{K}_{ij}$ with respect to $j$ for the massive case with $m=0.3$ and $L=80$ and after performing measurements on $10$ contiguous sites.
} 
\end{figure}
\begin{equation}\label{harmonicOsc}
\mathcal{H}=\frac{1}{2}\sum_{n=1}^{L}\pi_n^2+\frac{1}{2} \sum_{n,n^\prime=1}^{L}\phi_{n} K_{nn^\prime}\phi_{n^\prime}~.
\end{equation}
The ground state of the above Hamiltonian has the following form
\begin{equation}\label{GroundSwave}
\Psi_0=(\frac{\det K^{1/2}}{\pi^L})^{\frac{1}{4}} e^{-\frac{1}{2}\bra{\phi}K^{1/2}\ket{\phi}}.
\end{equation}
One can calculate the two point correlators $(X_D)_{ij}=\Tr(\rho_D \phi_i \phi_j)=(K^{-\frac{1}{2}})_{ij}$ and 
$(P_D)_{ij}=\Tr(\rho_D \pi_i \pi_j)=(K^{\frac{1}{2}})_{ij}$ using the $K$ matrix, where $\rho_D$ is the reduced density matrix of the domain $D$.
 The square root of this matrix, as well as its inverse, can be split up  into 
coordinates of the subsystems $D$  and $\bar{D}$ , 
i. e., 
\[ K^{-1/2}=\left( \begin{array}{cc}
X_{D} & X_{D\bar{D}} \\
  X^{ T}_{D\bar{D}} & X_{\bar{D}} 
  \end{array} \right),\hspace{1cm}
  K^{1/2}=\left( \begin{array}{cc}
P_{D} & P_{D\bar{D}} \\
  P^{ T}_{D\bar{D}} & P_{\bar{D}} 
  \end{array} \right)\]
The spectra of the  matrix $2C = \sqrt{X_DP_D}$, can be used to calculate              
 the von Neumann and R\'enyi entanglement entropies, 
 \begin{eqnarray}\label{Entanglement entropy1}
S=\tr\left[(C+\frac{1}{2})\ln(C+\frac{1}{2})-(C-\frac{1}{2})\ln(C-\frac{1}{2})\right],\\
\label{Entanglement entropy2}
 S_{\alpha}= \frac{1}{\alpha-1}\rm \tr\left[ \ln\left((C+\frac{1}{2})^{\alpha}-(C-\frac{1}{2})^{\alpha}\right) \right].\hspace{1cm}
\end{eqnarray}
 Now if we make a measurement on the position of all the oscillators $\{\phi_i\}\in A$ they will take some definite values and   
will get decoupled from the rest
of the oscillators. In other words, the post-measurement state will be the same as (\ref{GroundSwave}) but instead of $K^{1/2}$ we need to consider
$(K^{1/2})_{\bar{A}}$ which is a subblock of the matrix $K^{1/2}$ corresponding to the oscillators in the subsystem $\bar{A}$. This means that we now have a 
new Gaussian wave function and one can calculate its bipartite entanglement entropy
with the formulas (\ref{Entanglement entropy1}) and (\ref{Entanglement entropy2}). There is a 
simple interpretation for the wave function after measurement. It is just the ground state of
the Hamiltonian (\ref{harmonicOsc}) with $K=-\tilde{K}=((K^{1/2})_{\bar{A}})^2$. It is not difficult to see that this Hamiltonian is highly non-local,
for example, if we start with a discrete Laplacian in one dimension $K_{ij}=-\delta_{i,j-1}-\delta_{i,j+1}+2\delta_{i,j}$ the $\tilde{K}$ will have
the form shown in Figure ~2. Note that the $\tilde{K}_{ii}$ is usually a number very close to $-2$ and 
$\tilde{K}_{i,i+1}=\tilde{K}_{i+1,i}$ is very close to $1$ and the rest of the couplings are much smaller than these three couplings.
As it is shown in figure ~2 for site $i=1$ the couplings decrease like a power-law but after reaching the middle of the original 
system they start to increase. It seems  that this behavior is generic for all the oscillators up to the middle of the system.
The sites beyond the middle of the system just show reverse behavior. The behavior in the  massive case 
$K_{ij}=-\delta_{i,j-1}-\delta_{i,j+1}+(2+m^2)\delta_{i,j}$
is a bit different. In this case $\tilde{K}_{i,j}$ decreases exponentially as it is shown in Figure ~2. In this case the only significant
elements of the matrix $\tilde{K}$ are $\tilde{K}_{i,i+1}=\tilde{K}_{i+1,i}=1$ and $\tilde{K}_{ii}=-2-m^2$ which means that $\tilde{K}$ is actually
almost identical to the $K$ matrix of the local massive Laplacian. This result will have  significant consequences in the next sections when we discuss
post measurement entanglement entropy of massive systems. It is worth mentioning that we expect that the post-measurement wave function is the ground state
of a non-local Hamiltonian in generic systems. This makes the study of entanglement entropy in non-local Hamiltonians interesting
also from this perspective, for a list of studies on the entanglement entropy of non-local systems see 
\cite{violationlongrange,Eisert2006,Ghaseminezhad2012,Ghaseminezhad2013}. 
In the upcoming sections, we use the above two versions of couplings (massless and massive) as the discretized
versions of our free bosonic field theory.

\subsection{Numerical results for disconnected regions }

In this subsection we check the validity of the results presented in the section ~3 by numerical means. The equations
(\ref{EE subsreion}) and (\ref{EE subsreion finite size}) are the classical results that have been checked before by many methods \cite{Calabrese2009}.
The equation (\ref{SB for PBC}) is already checked numerically in \cite{Rajabpour2015b}.
Here we study the entanglement entropy of two regions completely decoupled from each other, in other words, we check the equations
(\ref{power-law decay}) and (\ref{power-law decay 2}). 
In $1+1$ dimensions the smallest scaling dimension in the Klein-Gordon field theory is $\Delta=1$ 
which is the one  corresponds to the field $\partial\phi$. Consequently we expect that 
the entanglement entropy decays like a power-law $(\frac{l}{s})^{\Delta(\alpha)}$ with an exponent which is $\alpha$ dependent up to $\alpha=1$ 
and then it saturates, see equation 
(\ref{power-law decay}). In other words the exponent will be
\begin{equation}\label{exponent}
{\Delta(\alpha)}=\left\{
\begin{array}{c l}      
    2\alpha, & \alpha<1\\
        2 &\alpha\geq1,\\        
\end{array}\right.
\end{equation}
The numerical
results depicted in the figures ~3 indeed confirm our expectations. We also checked the formula
(\ref{power-law decay 2}). For the Klein-Gordon field theory we should have power-law decay $(\frac{l}{L})^{\tilde{\Delta}(\alpha)}$ with
\begin{equation}\label{exponent2}
{\tilde{\Delta}(\alpha)}=\left\{
\begin{array}{c l}      
    4\alpha, & \alpha<1\\
        4 &\alpha\geq1,\\        
\end{array}\right.
\end{equation}
The results shown in figure ~4 is in a good agreement with the above formula.
\begin{figure} [htb] \label{fig3}
\centering
\includegraphics[width=0.4\textwidth,angle =-90]{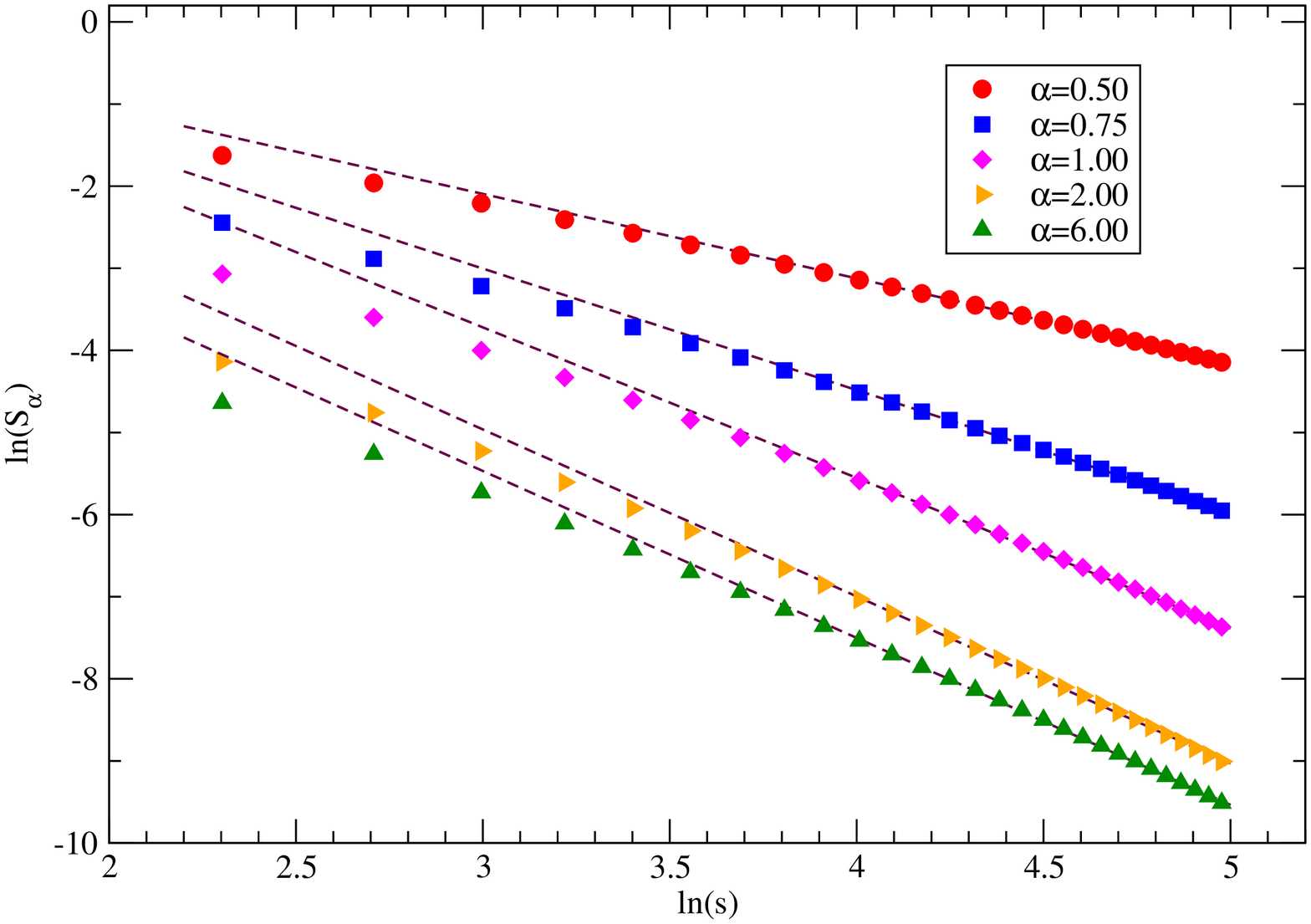}

\includegraphics[width=0.4\textwidth,angle =-90]{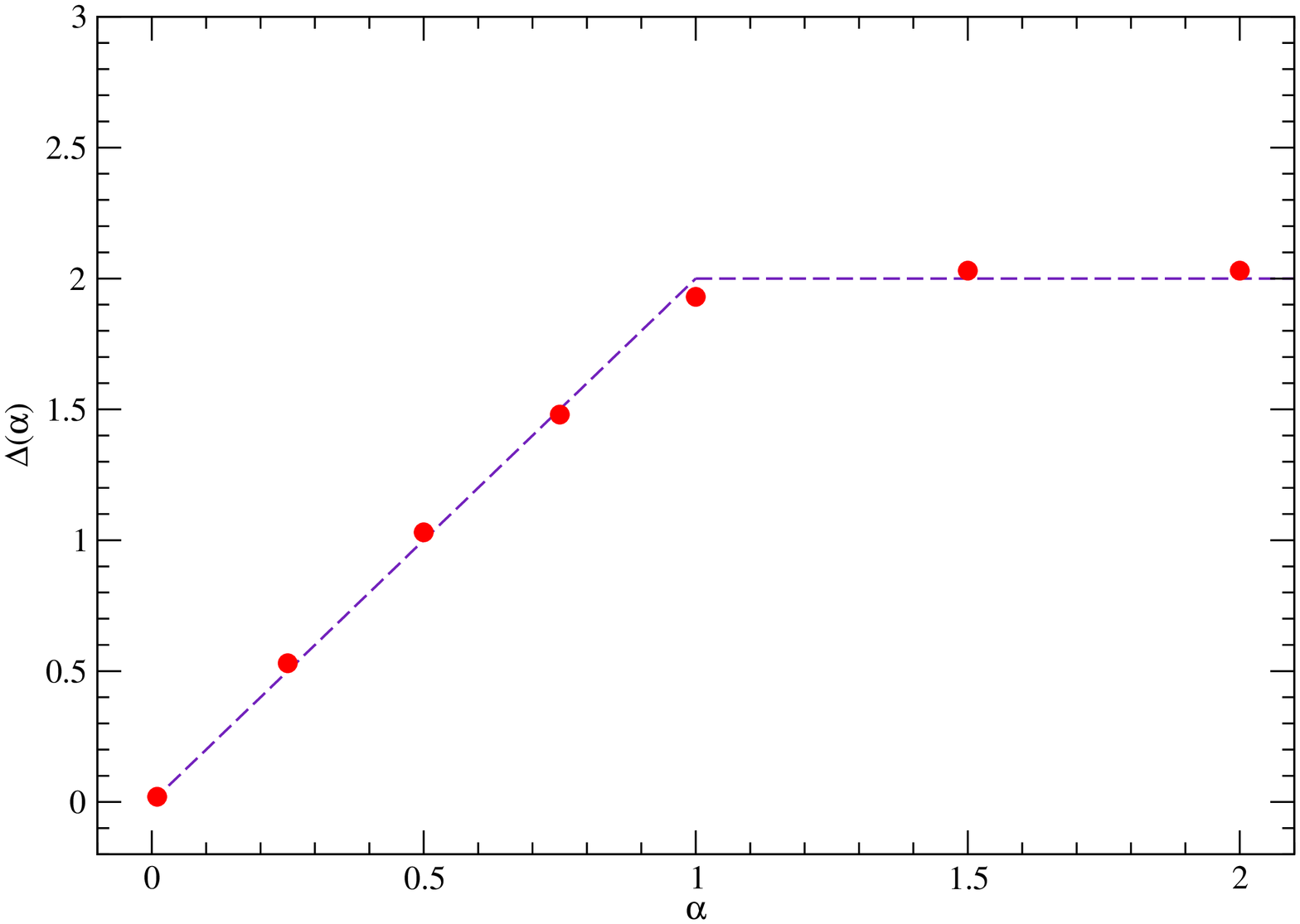}

\caption{(Color online) Top: log-log plot of the R\'enyi entropy for the two disconnected regions. From top to bottom 
the full lines correspond to equation (\ref{power-law decay})
with $\Delta(\alpha)$ coming from the equation (\ref{exponent}).
Down: the exponent of the power-law  i.e. $\Delta(\alpha)$, with respect to $\alpha$. The size of
the total periodic system  $L=1000$ and the size of  the region $B$ is $l=10$. The exponents are derived by fitting the data in the region
$s\in(80,150)$. The full line is the analytical result (\ref{exponent}).
} 
\end{figure}

\begin{figure} [htb] \label{fig4}
\centering
\includegraphics[width=0.4\textwidth,angle =-90]{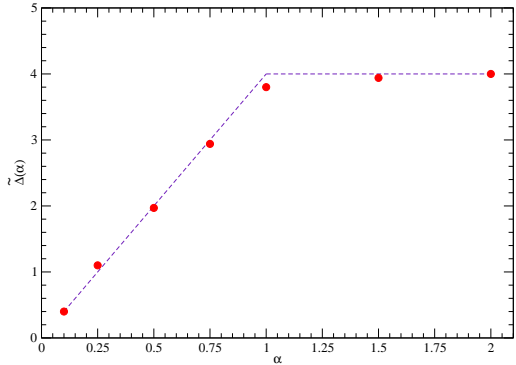}
\caption{(Color online) the exponent of the power-law  i.e. $\tilde{\Delta}(\alpha)$, with respect to $\alpha$. The size of
the total periodic system  $L=600$ and the size of  the regions $B$ and $\bar{B}$ is taken $l\in(20,50)$. 
The full line is the analytical result (\ref{exponent2}).}
\end{figure}

\subsection{$1+1$ dimensional massive QFT }

In this subsection we would like to investigate bipartite entanglement entropy after partial measurement in
$1+1$ dimensional massive Klein-Gordon field theory.

When the field theory is massive the bipartite entanglement entropy follows the area-law \cite{Casini2009a,Calabrese2009}. There are also many concrete proofs 
regarding the validity of the area-law in the $1+1$ dimensional gapped systems, see \cite{Hastings2007,Arad2012,Brandao,Arad2013}. In massive local
quantum field theories, it is shown \cite{Calabrese2009} that the entanglement entropy of the system is given by
\begin{eqnarray}\label{one d massive}
S_{\alpha}=-\kappa \frac{c}{12}(1+\frac{1}{\alpha})\ln m+...,
\end{eqnarray}
where $c$ is the central charge of the system which is equal to ~1 for Klein-Gordon field theory, $m$ is the mass and $\kappa>0$ is the number 
of contact points between two subsystems. We calculated the same quantity after
partial measurement, as we discussed before, for different cases. Interestingly we found that the equation (\ref{one d massive})
is valid also after we decouple the region $A$ from the system. One just needs to consider $\kappa$ as the number of contact points between 
$B$ and $\bar{B}$. We checked this for periodic boundary conditions for the cases that $B$ and $\bar{B}$ have one and two contact points. The results 
for $\alpha=1$ are shown in
the Figure ~5. \begin{figure} [htb] \label{fig5}
\centering
\includegraphics[width=0.4\textwidth,angle =-90]{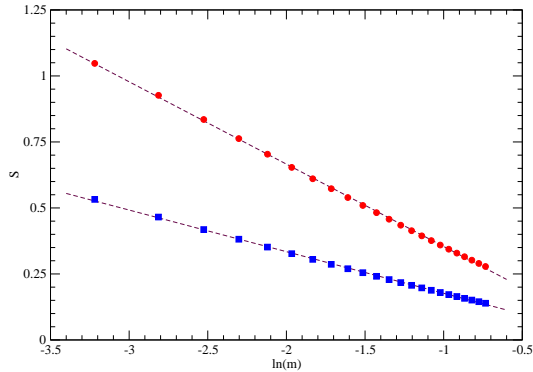}
\caption{(Color online) Post measurement von Neumann entanglement entropy   in massive field theory in $1+1$ dimension. The upper data correspond to 
when $B$ and $\bar{B}$ have two common boundary points and the lower data correspond to when $B$ and $\bar{B}$ have one
common boundary. The full  lines are the equation (\ref{one d massive}) with the coefficient of the logarithm from top to bottom 
being equal to
$-0.31$ and $-0.16$ respectively.  Mass range is chosen in a way that $1<\frac{1}{m}<L$, where in the above $L$ the size
of the whole system is $100$.}
\end{figure} Similar results are valid also for $\alpha>1$ confirming the equation (\ref{one d massive}). The above 
results look consistent with this picture that the entanglement between two regions is related to the couplings just around the 
interfaces of the two regions. Since as we discussed before the local measurements in  parts of the system which is far from the boundary
between $B$ and $\bar{B}$ disturb very little the couplings around the interfaces
one might find it natural to have still the area-law after the partial measurement. Although post measurement area-law is probably a generic behavior
of gaped systems the equation (\ref{one d massive}) is valid only in those cases in which we make our measurements in the conformal basis. We notice here
that since in our system the correlations $\Tr(\rho_D \pi_i \pi_j)$ and $\Tr(\rho_D \phi_i \phi_j)$ decay exponentially even after
the measurement, based on \cite{Brandao}, one naturally expects the area-law. This might be a good starting point for an analytical approach
to the problem in the most general basis.

We also calculated the entanglement entropy in the  case  $l\ll s_1=s_2=s$
 and found that  the entanglement entropy decays exponentially.
In other words we have
\begin{eqnarray}\label{one d separated 1 massive}
S_{\alpha}\asymp_{_{s\to\infty}} e^{-\gamma(\alpha) m s},	
\end{eqnarray}
where $\gamma(\alpha)$ is a  number that decreases with increasing $\alpha$.

\section{Conclusions}

In this paper we discussed the entanglement entropy of the ground state of the
 conformal field theories in the $1+1$ dimension after projective measurement in the conformal basis in part of the system.
Using the boundary conformal field theory methods we first rederived the known results of \cite{Holzey1994} and \cite{Rajabpour2015b}. Then we further showed that
if the two regions that we would like to calculate their entanglement  with respect to each other got spatially  decoupled
after the projective measurement
the entanglement entropy  decays like a power-law with respect to the distance with an exponent which is equal to $2\alpha\Delta$ for $\alpha\leq1$
and $2\Delta$ for $\alpha>1$, where $\Delta$ is the smallest scaling dimension in the spectrum of the system.  Similar
calculations are also done in the presence of the finite size effects. We then checked our results
using numerical calculation on the Klein-Gordon field theory.
We also provided numerical results regarding the massive case. In particular, we showed that after the projective measurement
 the bipartite entanglement entropy of the unmeasured oscillators follows the area-law 
as far as the two parts
are connected. When due to the measurement region they are disconnected the entanglement
entropy decays exponentially. It will be very interesting to check, analytically or numerically, the validity of  our results
in other $1+1$ dimensional systems such as spin chains.
Calculating the same quantities in higher dimensions is also stimulating.
Finally, it will be nice  to investigate the same kind of questions using holographic techniques  \cite{Ryu}.

\paragraph*{Acknowledgments.} We  are indebted to Pasquale Calabrese for many discussions and also
mentioning a crucial mistake in the early version of the manuscript. We also thank K. Najafi for numerous helps in numerical calculations. This work was supported in part by
CNPq.
\newline
\newline

\appendix 
\section*{Appendix: Conformal maps}
\setcounter{section}{1}
In this appendix we provide some details regarding the conformal maps used in the main text. The conformal map from a whole plane 
with two slits and a branch cut to annulus can be derived as follows: the first step is mapping the whole plane with two slits to a whole plane
with two symmetric slits as it is shown in the Figure ~1. This can be done with the map $\tilde{z}(z)$ which has the following form:
\begin{eqnarray}\label{tilde z}
\tilde{z}(z)&=&\frac{2a}{k}\frac{z}{bz+1}-\frac{1}{k},\nonumber\\
a&=&\sqrt{\frac{s_2(s_2+l)}{s_1(s_1+l)}}\frac{1}{l+s_1+s_2},\nonumber\\
b&=&\frac{\sqrt{s_1s_2(l+s_1)(l+s_2)}-s_2(l+s_1)}{(l+s_1)(s_1s_2-\sqrt{s_1s_2(l+s_1)(l+s_2)})},\nonumber\\
\end{eqnarray}
with the parameter $k$ given by
\begin{eqnarray}\label{k}
k=1+2\frac{s_1s_2-\sqrt{s_1s_2(l+s_1)(l+s_2)}}{l(l+s_1+s_2)}.
\end{eqnarray}
Then the whole plane with the symmetric slits and a branch cut can be mapped to annulus with a branch cut by the following conformal map \cite{Kober}:
\begin{eqnarray}\label{conformal map annulus}
w_{1}(\tilde{z})&=&e^{-\frac{h}{2}}e^{h \frac{\mbox{sn}^{-1}(\tilde{z},k^2)}{2\mathcal{K}(k^2)}},\\
h&=&2\pi\frac{\mathcal{K}(k^2)}{\mathcal{K}(1-k^2)},
\end{eqnarray}
The final step is the uniformization of the domain by using the conformal map $w_{\alpha}(z)=w_{1}(z)^{\frac{1}{\alpha}}$.

\begin{figure} [htb] \label{fig1}
\center
\includegraphics[width=0.8\textwidth]{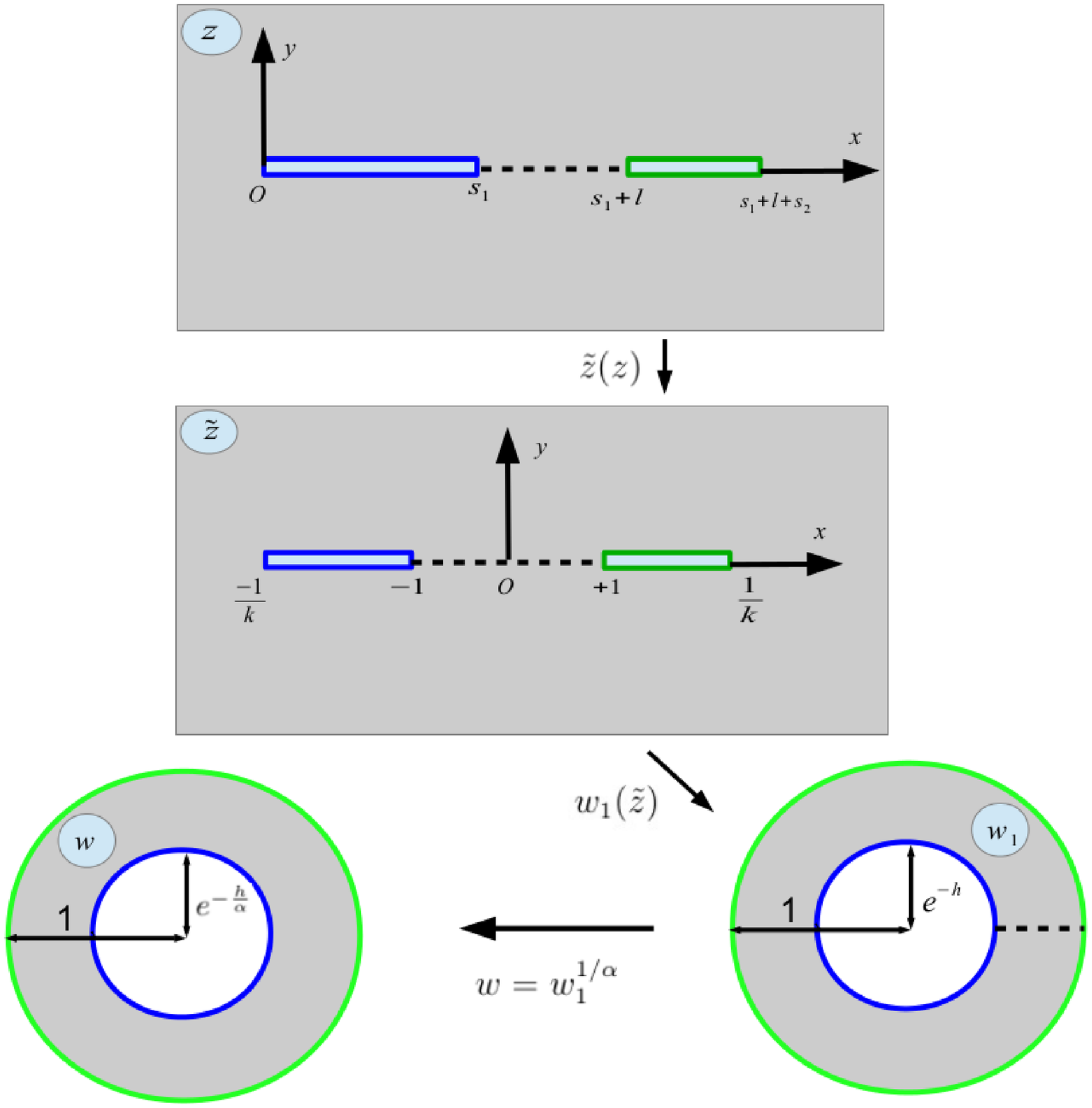}
\caption{(Color online) Mapping between different regions. The whole plane with two slits (blue and green) and branch cut (dashed line)  can be mapped to
the whole plane with symmetric slits by the map $\tilde{z}$. Using $w_{1}(\tilde{z})$ we can map the whole plane with symmetric slits and a branch cut to
annulus with a branch cut. Finally $w_{1}(z)^{\frac{1}{\alpha}}$ removes the branch cut.} 
\end{figure}
The same procedure as above can be used also for the periodic system. We first use the conformal map $e^{2\pi i \frac{z}{L}}$ to map the cylinder
with two slits and a branch cut
to the whole plane with slits and a branch cut. Then we map the remaining region to a whole plane with symmetric slits by using a mobius map.
The above two steps can be summarized as the equation (\ref{conformal map2 details}). Now the whole plane with symmetric slits and a branch cut
can be mapped to an annulus by the map $\Big{(}w_{1}(\tilde{z})\Big{)}^{\frac{1}{\alpha}}$.

\end{document}